\documentclass[aps,prl,twocolumn,superscriptaddress]{revtex4}
\usepackage{amsmath}
\usepackage{amssymb}
\usepackage{graphicx}
\usepackage{footnote}
\usepackage{subfigure}
\usepackage{sidecap}
\usepackage{verbatim} 
\usepackage{microtype}
\usepackage{hyperref}
\usepackage{float}
\usepackage{bm}
\usepackage{setspace}

\begin{document}

\title{Dynamics and Scission of Rodlike Cationic Surfactant Micelles in Shear Flow}
\author{Abhinanden Sambasivam}
\affiliation{Department of Biomedical and Chemical Engineering, Syracuse University, Syracuse, NY 13244}
\author{Ashish V. Sangwai}
\affiliation{Intel Corporation, Hillsboro, OR 97124}
\author{Radhakrishna Sureshkumar}
\email{rsureshk@syr.edu }
\affiliation{Department of Biomedical and Chemical Engineering, Syracuse University, Syracuse, NY 13244}
\affiliation{Department of Physics, Syracuse University, Syracuse, New York 13244, USA}
\date{\today}

\begin{abstract}

Flow-induced configuration dynamics and scission of rodlike micelles are studied for the first time using molecular dynamics simulations in presence of explicit solvent and salt. Predicted dependence of tumbling frequency and orientation distribution on shear rate $S$ agrees with mesoscopic theories. However, micelle stretching increases the distance between the cationic head groups and adsorbed counter ions, which reduces electrostatic screening and increases the overall energy $\Phi$ linearly with micelle length. Micelle scission occurs when $\Phi$ exceeds a threshold value, independent of $S$.
\end{abstract}

\maketitle
Thermodynamic self-assembly in surfactant solutions is known to result in various micelle morphologies such as spheres, cylinders, vesicles, and lamellae. The rich diversity in equilibrium morphologies is of great fundamental and practical interest in applications ranging from detergency to targeted drug delivery. Amongst micellar structures, cylindrical and wormlike micelles are known to exhibit rich rheological behavior because micelle length and entanglement density depend greatly on the surfactant and salt concentration, salt hydrophobicity, temperature and flow shear. Specifically, depending on the surfactant concentration, chemical environment, and shear rate, cylindrical micelle solutions could exhibit Newtonian, shear thinning, or shear thickening behavior, which is accompanied by shear-induced structure (SIS) formation or shear banding \cite{Rehage86, Wunderlich87, Cates92, Mukund10, Mukund08, Pine96,Miller07}. While phenomenological, kinetic-theory inspired \cite{Vasquez07} and continuum-level constitutive models \cite{Manera03, Morozov12, Fielding10, Fielding04} have been developed to study the dynamics and rheology of cylindrical micelles, molecular-level explorations of flow-structure coupling in micelle solutions is lacking. Understanding micelle structure and dynamics via faithful and systematic molecular simulations is an important and essential step in filling this knowledge gap \cite{Dhakal14}. Briels et al. \cite{Padding05} have developed stochastic simulations (Brownian dynamics), in which parameters such as the persistence length were obtained from atomistic molecular simulations, to study dynamics and rheology of wormlike micelle solutions. Here, we describe for the first time insights gained from a coarse-grained molecular dynamics (CGMD) study of cylindrical cationic micelles subjected to uniform and steady shear flow. In CGMD, explicit solvent-salt-micelle interactions are incorporated by utilizing force fields which are validated against more detailed atomistic MD simulations \cite{Sangwai11, Sangwai12}. Several fundamental questions are addressed including the criterion that demarcates the Brownian (diffusive) and flow-aligned regimes, asymptotical scaling laws that characterize micelle configurational dynamics and their comparisons with stochastic theories of flexible and semi-flexible polymers \cite{Chu99, Chu98, Chu05, Winkler06}, relationship between micelle energetics and length, and a mechanism of flow-induced micelle scission.

The configurational dynamics characterized by quasi-periodic cycles of flow-alignment, micelle extension and tumbling dictate inter-micelle interactions, which under favorable conditions can cause merger of cylindrical micelles to form larger SISs \cite{Rehage86, Wunderlich87,Mukund10, Mukund08, Pine96, Miller07}. The critical shear rate required to induce SIS formation is believed to be strongly correlated to that required to cause significant flow-alignment of the micelle. Cates and coworkers \cite{Cates92} suggested that the fundamental mechanism of SIS formation is end-to-end collisions between flow-aligned micelles, followed by the opening up of micelle end caps and micelle fusion. They hypothesized that the onset of SIS formation occurs when the rotational (angular) diffusion timescale $\tau_d$ of the micelles is comparable to the inverse shear rate, i.e., when the Peclet number $Pe$$\equiv$$S$$\tau_d$, is O(1). Recent equilibrium MD simulations of cationic spherical micelles suggest that sufficient electrostatic screening is required to induce micelle fusion, which is more readily facilitated when ringed organic counter ions are present in salt to surfactant molar ratio exceeding unity (see Fig. 5 in \cite{Sangwai12}). For instance, benzoate ions physically adsorb onto the micelle-water interface and help neutralize the repulsive electrostatic interactions between the surfactant head groups, as characterized by the potential of mean force of binary interactions. This is consistent with experimental observations that hydrophobic salts such as benzoate or salicylate promote SIS formation and shear thickening in cationic micelle solutions \cite{Oelschlaeger10}.

Hence, the literature suggests that satisfaction of two criteria is required for robustly inducing SIS in ionic micelle solutions, namely a mechanical one that requires that the inverse rate of flow deformation is comparable to that of the inherent rotational time scale of the micelle, and a chemical one that necessitates the counterion concentration to be large enough to provide sufficient electrostatic screening. In this work, we focus on studying the effects of shear flow on the configurational dynamics of a single cylindrical Cetyl trimethylammonium chloride (CTAC) micelle in an aqueous solution containing sodium salicylate (NaSal) salt while preserving the underlying physical chemistry and intrinsic self-assembled structure of the micelle. In the simulations, all chemical interactions, spontaneous rearrangements within the micelle structure and micelle shape/length fluctuations and angular diffusion are explicitly calculated. To date, atomistic non-equilibrium molecular dynamics (NEMD) simulations of micellar systems in presence of explicit solvent and electrostatic interactions are limited to ns time scales. In the present study, we are able to conduct NEMD studies that span microseconds ($\gg$ angular orientational relaxation time $\tau_r$) by utilizing coarse-grained (CG) potentials of CTAC surfactant and NaSal salt in explicit solvent. Coarse-graining of surfactant, salt, and water helps reduce the number of particles by approximately 4 times and increase the time step in MD by an order of magnitude since time scales on the order of atomic bond vibrations are neglected in the CG description. MARTINI-based CG force fields \cite{Marrink07} used here have been extensively validated against atomistic simulations in terms of the radial distribution functions of the system constituents \cite{Sangwai11}. Polarizable CG water was used as a solvent because of its ability to capture electrostatic interactions in systems with charged interfaces more accurately compared to CG water models that lack a local dipole \cite{Marrink10}. The integrity of the cylindrical structure of the micelle is maintained by the surface adsorption of Sal$^-$ ions onto the micelle-water interface, especially when the molar ratio of NaSal to CTAC, $R$$\geq$$1$. Hence in the simulations we used $R$$=$$1.5$. The methodology of assembling and equilibrating cylindrical micelle structure, implementation of shear flow and boundary conditions in the MD simulations, determination of micelle relaxation from flow-induced stretched configurations, and the calculation of the instantaneous micelle length and orientation are provided in the Supplemental Information \cite{Suppinf01}. The Weissenberg number $Wi\equiv$S$\tau_r$ is used to characterize the flow strength. Simulations are conducted for $0$$\leq$$Wi$$\leq$$125$. The vector that connects the centers of mass of the two endcaps of the micelle represents the micelle axis. The angle $\theta$ that the projection of this vector onto the flow (x)-gradient (y) plane subtends with the x-direction is used to quantify orientation and flow-alignment. Micelle structure is not subject to any external constraints and is governed only by hydrodynamic forces as well as the interactions among the CTAC surfactants, sodium salicylate salt (NaSal), Cl$^-$ counterions, and water molecules.

\begin{figure}
\includegraphics[width=3.2in]{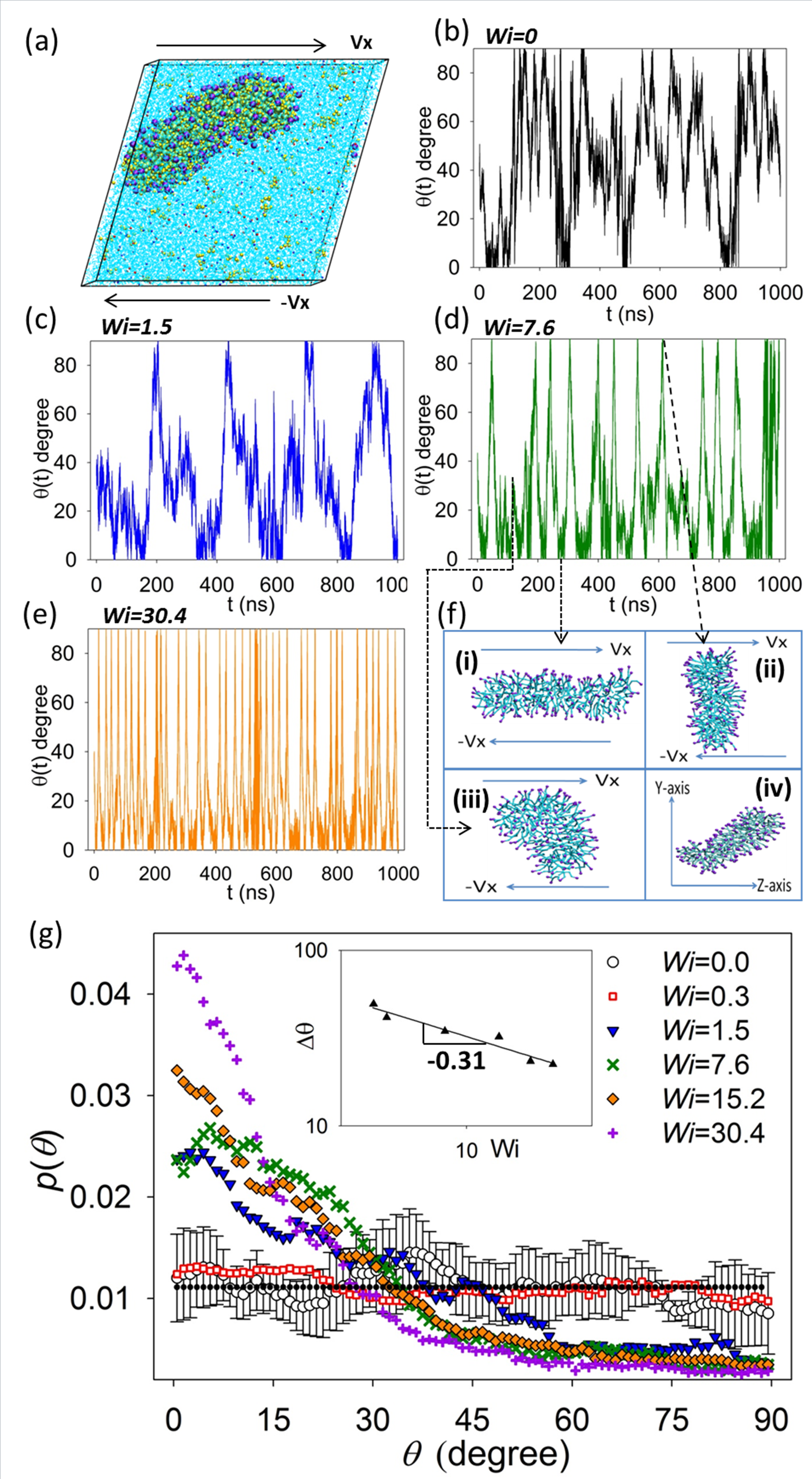}
\caption{ \label{figure1}(Color online) (a) Schematic of NEMD simulations.  CTA$^+$ surfactants (purple head and green tails), Sal$^-$ ions (yellow), Na$^+$ (red) and Cl$^-$ (blue) ions and water (cyan). (b)-(e): Plots of $\theta$ versus t at for various $Wi$. (f): (i) Flow-aligned, (ii) gradient-aligned, (iii) misaligned (in the xy plane), and (iv) misaligned (in the yz plane) states. (g): Probability distribution function p($\theta$) at various $Wi$. The dotted line represents a uniform distribution and the vertical bars show the standard deviation of p($\theta$) at $Wi$$=$0 in the simulations within each interval. Inset shows a plot of $\Delta\theta$ as a function of $Wi$ showing $\Delta\theta$$\propto$$Wi$$^{-0.31}$}
\end{figure}

Figure \ref{figure1}(a) shows a schematic of the NEMD simulation box and Fig. 1(b-e) shows $\theta$ plotted as a function of time $t$ for 4 different values of $Wi$. Note that $\theta$$=$0$^{\circ}$ represents a fully flow-aligned state while a transition through 90$^{\circ}$ corresponds to a tumbling event. Results are shown only for the first 1 $\mu$s for clarity although the trends remain the same throughout the duration of the simulations (3$\mu$s). It is evident that in the absence of flow, i.e., for $Wi$=0, the intrinsic rotational diffusion of the micelle dominates the orientational distribution. This is clearly seen in Fig. 1(g) where the probability distribution function (pdf) p($\theta$) is plotted. The pdf is uniform for $Wi$$=$$0$ suggesting the absence of any preferred orientation. As $Wi$ is increased, the effect of flow on micelle orientation becomes progressively more prominent. For example, for $Wi$$=$$0.3$ the departure of p($\theta$) from the  uniform equilibrium distribution is relatively small suggesting that the rotational diffusion dominates over flow alignment. However, at $Wi$$=$$1.5$, higher probabilities are clearly seen associated with flow-aligned states. A crossover from a diffusive to a flow-aligned regime occurs at $Wi$$\sim$$1$. For $Wi$$>$$1$, quasi-periodic tumbling events dominate the micelle dynamics. Snapshots (i) and (ii) of the micelle displayed in Fig. 1(f) correspond to $\theta$$=$0$^{\circ}$ (flow-aligned state) and $\theta$$=$90$^{\circ}$ (gradient-aligned state) respectively at $Wi$$=$$7.6$. The tumbling events occur at regular intervals while the tumbling frequency increases with $Wi$. For $Wi$$>$$1$, p($\theta$) follows a Gaussian distribution and the width at half maximum of the orientational distribution ($\Delta\theta$)$\propto$$Wi$$^{-0.31}$, which is consistent with previous predictions of mesoscopic theories for semiflexible polymers \cite{Winkler06, Winkler11, LeeKim09}. Shape fluctuations in the z-dimension also influence the tumbling cycles, as seen in snapshot (iii) in Fig. 1(f), in which the micelle is slightly bent and misaligned with respect to the flow direction (more snapshots are shown in \cite{Suppinf01}). In this case, the micelle axis is out of the flow-gradient plane, as seen in snapshot (iv). Such events occur more frequently for relatively low shear rates and tend to randomize the tumbling cycles. The micelle eventually recovers from such misaligned states to flow-aligned ones before undergoing a tumbling event or transitioning to another misaligned state.

\begin{figure}
\includegraphics[width=3.00in]{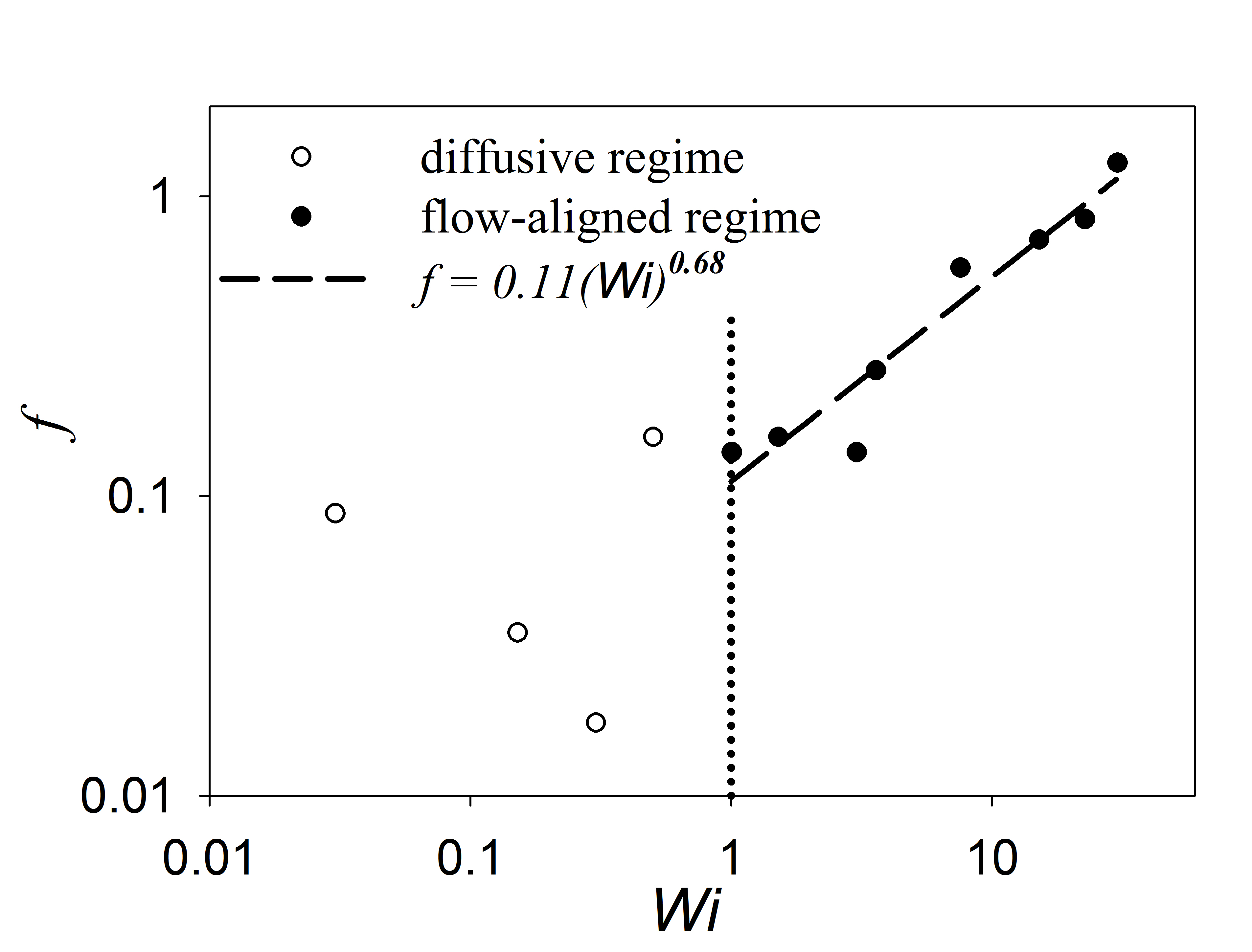}
\caption{ \label{figure2} Normalized tumbling frequency $f$ versus $Wi$. Open circles indicate stochastic diffusion regime while filled circles lie in the flow-aligned regime. Power law fitting in the flow-aligned regime yields $f$$\sim$$Wi$$^{0.68}$.}
\end{figure}

The autocorrelation function (ACF) of $\theta$ defines the temporal correlation of the angular orientation. The ACF at $Wi$$=$$0$ (at equilibrium) can be used to define an orientational relaxation time $\tau_r$ of $\theta$ by fitting it to a decaying exponential of the form, 
\begin{equation}
ACF=\exp\left(-t/\tau_r \right),
\end{equation}
where the ACF has been normalized using its value at $t$$=$$0$. The relaxation time $\tau_r$ was calculated to be 28.7 ns and was used in our calculations of $Wi$.  The power spectral density (PSD) can be obtained from the fast Fourier transform (FFT) of the ACF. We obtain the tumbling frequency $f^*$$\equiv$$(1 / \tau_c)$ from the peak in the PSD \cite{Schroeder05}. The normalized tumbling frequency $f$$\equiv$$\frac{1}{[(\tau_c / \tau_r)]}$ is plotted as a function of $Wi$ is shown in Fig. \ref{figure2}. For $Wi$$<$$1$,  no systematic change in $f$ is observed. However, for $Wi$$>$$1$, $f$ increases indicating more frequent and periodic tumbling. Power law fitting of $f$ vs. $Wi$ provides an exponent of 0.68 which is also observed in Brownian dynamics simulations of polymers in dilute solutions \cite{Celani05,Chertkov05,Puliafito05,Sultan10}. The rotational diffusivity D was obtained from equilibrium simulations as one half of the slope of the linear fit of the mean squared angular displacement vs. time. This yielded $D=6\times10^6$rad$^2$s$^{-1}$, corresponding to $\tau_D$$=$6.58$\times$$10^{-6}$s. In the transition region ($Wi$$=$$1$) $Pe$$\approx$230. Hence, we find $Wi$$\sim$$1$ to be a more appropriate criterion for inducing flow-alignment, suggesting that flow-alignment occurs when the orientational relaxation time of the micelle is comparable to the inverse shear rate.

\begin{figure}
\includegraphics[width=3.0in]{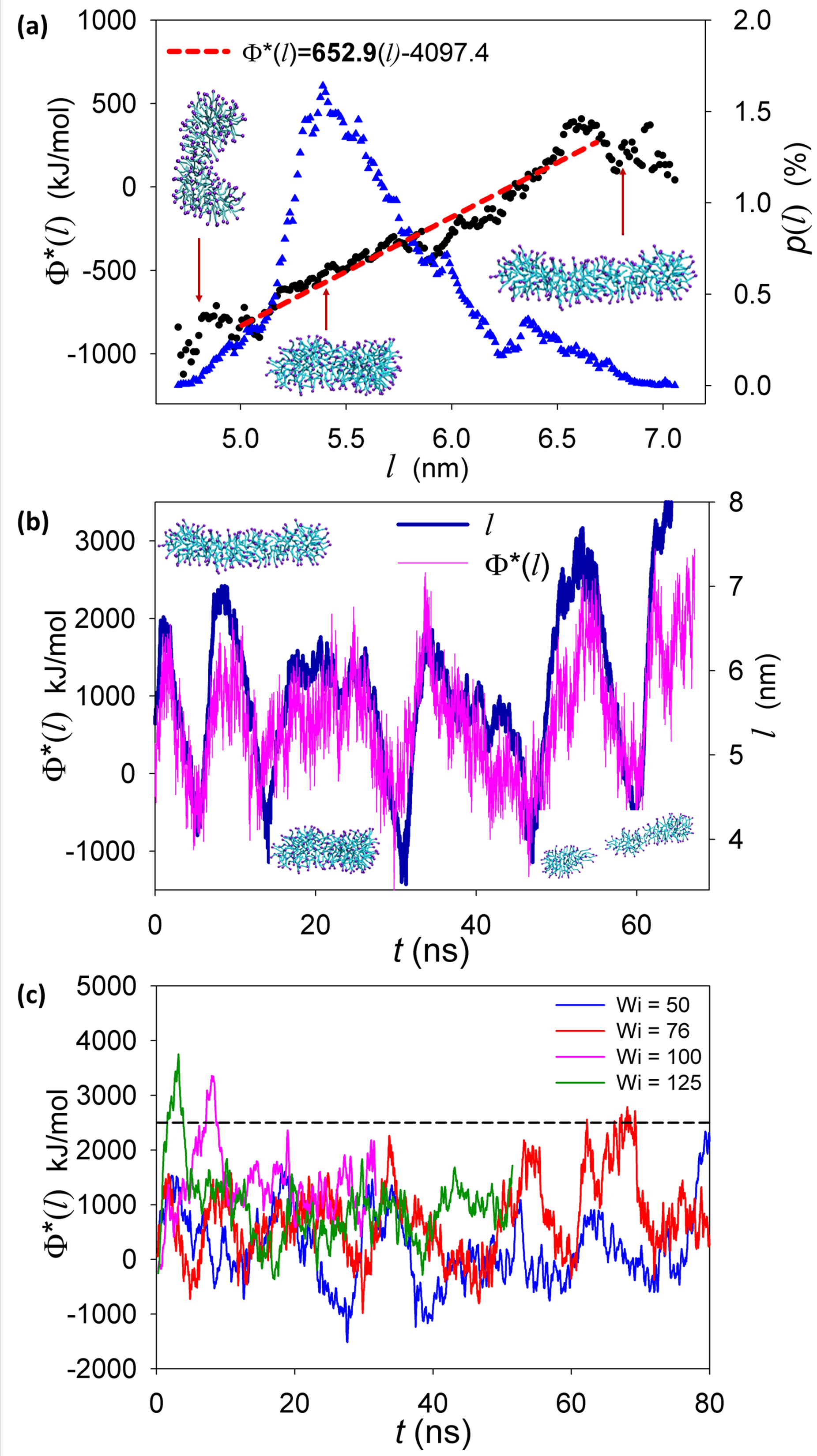}
\caption{ \label{figure3}(Color online) (a): Pair potential energy $\Phi^*(l)$ (black dots) and probability distribution of length $p(l)$ (blue triangles) from the length relaxation simulations are plotted as a function of $l$. (b) $\Phi^*(l)$ and $l$ versus $t$ showing micelle scission. Snapshots at different locations show 1. stretched micelle, 2. micelle with most probable $l$, and 3. micelle after scission. (c)  $\Phi^*(l)$ vs. $t$ for 4 different simulations indicating a $Wi$-independent threshold for micelle scission for $Wi$$>$70.}
\end{figure}

We computed the change in the total pair potential energy $\Phi(l)$ of the surfactants in the micelle and adsorbed Sal$^-$ ions as a function of length $l$ of the micelle. Length $l$ is the magnitude of the vector $\textbf{l}_{axis}$ connecting the centers of mass of the two end-caps of the micelle. Note that $l$ therefore is smaller than the end-to-end distance of the micelle. The use of $\textbf{l}_{axis}$ to calculate $l$ instead of using the end-to-end micelle vector has an advantage of reducing the statistical uncertainty in $l$ arising from shape fluctuations. We carried out 10 different equilibrium MD simulations of previously sheared systems where flow-aligned stretched micelles were allowed to relax to their equilibrium length structures. Over the length relaxation spectrum of the micelle from about $l$=7.2 to 4.7nm, $\Phi^*(l)\equiv\Phi(l)-$$<$$\Phi(l, Wi=0)$$>$ was averaged into equally spaced bins of $l$ with a width d$l$=0.0122 nm (Fig. 3(a)). The relaxation of length happens over a period of roughly 8 ns and therefore only the first 8 ns of data was considered. Energy $\Phi^*(l)$ needed to stretch a micelle from its equilibrium length, calculated as the summation of pair potentials, is observed to vary linearly with $l$ (Fig. \ref{figure3}(a)). The slope of the linear fit provides a constant stretching force of $\approx$650 kJmol$^{-1}$nm$^{-1}$. In comparison, the force required to stretch a covalent bond is 2-3 orders of magnitude greater \cite{Jorgensen96}, while that for hydrogen bond stretching in water is 1-2 orders of magnitude greater \cite{Cremer12}. This is reasonable considering that the micelle is self-assembled by weak van der Waals forces. 

The linear relationship between $\Phi^*(l)$ vs. $l$ may be interpreted as an asymptotic behavior valid for a relatively short ($\sim$nm), stiff rodlike molecular assembly. From a further coarse grained, mesoscopic modeling point of view, this would suggest that a micelle with length on the order of a few nanometers may be considered as a stiff spring whose length may only fluctuate within a small range around an average value, e.g. as in the stiff FENE-Fraenkel spring proposed by Hsieh et al. \cite{Larson06} to model practically inextensible rods in the freely jointed bead-rod chain model of polymer chains. In order to verify this, we calculated the probability distribution of $l$ from a total of 160,000 configurations. A probability distribution of $l$ during the relaxation process is also plotted in Fig. 3(a). The distribution shows a pronounced peak at 5.4 nm which is the most probable length. The micelle appears to explore lengths smaller than $l$=5.4 nm which is possible because of a mechanism in which the recoil from a stretched state to equilibrium length would cause either micelle bending or compression. In the three different micelle states shown in the plot, a stretched micelle ($l$$\approx$7.0 nm) can be seen to relax to a shorter state with $l$$\approx$5.4 nm and then further recoil into a state bent along its axis ($l$$<$5.0 nm).

Unlike polymer chains, in which the monomers are covalently bonded to each other, micelles are self-assembled by weak non-bonded interactions. This presents a limitation in the range of $Wi$ that can be explored without causing micelle scission. We performed multiple simulations between $50$$\leq$$Wi$$\leq$$125$ to study micelle scission. It was observed that micelle scission does not occur for $Wi$$<$$70$. For larger $Wi$, flow shear is sufficiently strong to cause micelle breakage as shown in Fig. 3(b), in which the micelle is seen to undergo a few tumbling events before scission. A plot of $\Phi^*(l)$ as a function of time is shown in Fig. 3(c) for a case where no breakage was observed ($Wi$$=$$50$) and for three cases for which the micelle broke apart into two shorter ones ($Wi$$=$76,100,125). This plot, as well as data from other simulations, suggest the existence of an energy threshold for scission. From a series of simulations, the threshold value was estimated to be $\sim$2500-2700kJmol$^{-1}$. The observed extension before breakage, compared to the equilibrium configuration, is 2-3 nm. This, based on the stretching force estimated above, corresponds to an increase in $\Phi^*(l)$ of 1300-1950kJmol$^{-1}$, which is comparable to the estimated threshold. As the micelle stretches, the distance between the CTA$^+$ head group and adsorbed Sal$^-$ ions increases thereby reducing electrostatic screening and increasing the overall micelle energy. Above the threshold, it is energetically favorable to have two smaller micelles. This is further evidenced by the fact that the magnitudes of $\Phi^*(l)$ and $\Phi^*_{EL}(l)$, which is the electrostatic component of the pair potential energy between  CTA$^+$ and Sal$^-$, are approximately equal. The other components of $\Phi^*(l)$ remain practically unchanged.

In summary, we have provided a quantitative description of the shear-induced orientation dynamics, stretching and scission of rodlike surfactant micelles. An appropriate parameter that demarcates the diffusive and flow aligned regimes is the ratio of the micelle angular relaxation time to the inverse shear rate. Analyses of tumbling frequencies, orientation distributions and energy-extension relationships suggest that from a mesoscopic point of view, short rodlike micelles may be represented by a stiff bead-spring unit subject to hydrodynamic drag, Brownian fluctuations and a constant entropic spring force. Micelle scission happens through a mechanism in which shear-induced stretching causes the surfactant head groups and adsorbed counterions to be farther apart, resulting in reduced electrostatic screening and an increase in the overall micelle energy. The CGMD methodology used here can be extended to solutions that contain multiple micelles and/or nanoparticles to study emerging morphologies, flow-structure interactions and rheological behaviors to benefit applications ranging from enhanced oil recovery and hydrofracking \cite{Suliemanov11} to design of active nanofluids for energy harvesting and sensing \cite{Tao11}.

\acknowledgments
The authors gratefully acknowledge National Science Foundation grant CBET-1049454, CBET-1049489, computational resources provided by the Extreme Science and Engineering Discovery Environment (XSEDE), and Ms. Yutian Yang for input on tumbling frequency calculations.

\end{document}